\documentclass[12pt]{article}

\def\red{
}
\def\black{
}

\usepackage{epsfig}%

\usepackage[perpage,symbol*]{footmisc}

\def\addcite#1{[???]}
\def\chisb{\raise0.2em\hbox{$\chi$}SB}
\def\chiSM{\raise0.17em\hbox{$\chi$}SM}

\def\gray{\special{ps: 0.4 setgray}}
\def\black{\special{ps: 0.0 setgray}}

\newcommand{\draft}{
\newcount\timecount
\newcount\hours \newcount\minutes  \newcount\temp \newcount\pmhours

\hours = \time
\divide\hours by 60
\temp = \hours
\multiply\temp by 60
\minutes = \time
\advance\minutes by -\temp
\def\hour{\the\hours}
\def\minute{\ifnum\minutes<10 0\the\minutes
            \else\the\minutes\fi}
\def\clock{
\ifnum\hours=0 12:\minute\ AM
\else\ifnum\hours<12 \hour:\minute\ AM
      \else\ifnum\hours=12 12:\minute\ PM
            \else\ifnum\hours>12
                 \pmhours=\hours
                 \advance\pmhours by -12
                 \the\pmhours:\minute\ PM
                 \fi
            \fi
      \fi
\fi
}
\def\fullclock{\hour:\minute}
\gray
\font\Hugett  =cmtt12 scaled\magstep2
{\Hugett \strut \kern-3em Draft: \today,\clock}
\black
} 
\catcode`\@=11 
\def\lsim{\mathrel{\mathpalette\@versim<}}
\def\gsim{\mathrel{\mathpalette\@versim>}}
\def\@versim#1#2{\vcenter{\offinterlineskip
        \ialign{$\m@th#1\hfil##\hfil$\crcr#2\crcr\sim\crcr } }}
\catcode`\@=12 

\def\gappeq{\gsim}

\def\nextline{\hfill\break}

\def\mycomm#1{\nextline\strut\kern-6em{\red\tt ====> \ #1\black}\nextline}
\def\nextline{\hfill\break}
\newcommand{\beq}{\begin{equation}}
\newcommand{\eeq}{\end{equation}}
\newcommand{\bea}{\begin{eqnarray}}
\newcommand{\eea}{\end{eqnarray}}
\newcommand{\Seff}{S_{\kern-0.1em\hbox{\small \it eff}}}
\newcommand{\Leff}{{\cal L}_{\kern-0.1em\hbox{\small \it eff}}}
\newcommand{\SMB}{S_{\kern-0.1em\hbox{\small \it MB}}}
\newcommand{\Smb}{S_{\kern-0.1em\hbox{\small \it m-b}}}

\def\eff{\hbox{\small\it eff}\,}
\def\SeffU{S_{\kern -0.1em \eff}[u]}

\def\eqref#1{(\ref{#1})}
\def\naive{na\"{\i}ve}

\def\upstrut{\vrule height 1.1em depth 0.0ex width 0pt}

\begin{document}
\begin{flushright}
CERN-PH-TH/2005-001\\
TAUP--2792-05\\
Cavendish-HEP-05/02\\
\end{flushright}
\begin{center}
{\Large\bf Direct Estimate of the Gluon Polarization in 
the Nucleon\upstrut}
\end{center}
\vskip1cm
\begin{center}
{\bf John Ellis}\footnote{\tt john.ellis@cern.ch}\\
{\em Theory Division, CERN, Geneva, Switzerland}\\
\vspace*{0.2cm}
{\bf Marek Karliner}\footnote{\tt marek@proton.tau.ac.il}\\
{\em Cavendish Laboratory,
Cambridge University, England}\\
{\em \small on leave from}\\
{\em School of Physics and Astronomy\\
Raymond and Beverly Sackler Faculty of Exact Sciences\\
Tel Aviv University, Tel Aviv, Israel}\\
\end{center}
\vskip0.5cm

%
\renewcommand{\baselinestretch}{0.95}
\centerline{\bf Abstract}
\vspace*{5mm}
\noindent
We make a first crude direct estimate of the net gluon polarization in
the proton, $\Delta G$, combining data on the asymmetries in high-$p_T$
hadron production from the HERMES, SMC and COMPASS collaborations.
Although these data sample a restricted range of $x$, they provide no hint
that $\Delta G$ is large. Fixing the
normalizations of different theoretical parametrizations using the
hadron asymmetry data, we find typical central values of $\Delta G \sim
0.5$, with uncertainties of similar magnitude. Values of $\Delta G \ge
2$ are disfavoured by $\Delta \chi^2 \sim 9$ to 20, depending on the
parametrization used.
\vspace*{0.5cm}
\begin{flushleft} CERN-PH-TH/2005-001\\
V2, July 2005
\end{flushleft}


\vfill\eject
\section{Introduction}

The spin structure of the proton is still uncertain, despite
considerable experimental effort, theoretical ingenuity and several
surprises~\cite{g1review}. It certainly differs from \naive\ expectations
formulated
within the non-relativistic quark model, with strange quarks
apparently polarized oppositely to the proton: $\Delta s <
0$~\cite{Mallot}, and the
quarks altogether apparently contributing only about 30\% of the proton
spin: $\Delta \Sigma \equiv \Delta u + \Delta d + \Delta s \sim 0.3 \pm
0.1$~\cite{Mallot}. However, whether the remainder of the proton spin is
due to
gluons $\Delta G$ and/or orbital angular momentum $L_z$ remains an open
question~\cite{g1review}. One particular theoretical interpretation of the
proton spin
is provided by chiral soliton models, which suggest that $\Delta
\Sigma, \Delta G \sim 0$, with $L_z \sim 0.5$, up to corrections of
higher order in $1/N_c$ and $m_s$ \cite{BEKEK}.

The question of the magnitude of $\Delta G$ was given a high profile
by suggestions that gluons might be making a significant negative
contribution to the net quark spins,
\cite{Efremov:1988zh}-\cite{Carlitz:1988ab}
 which might even be sufficient to
explain all the negative polarization of the strange quarks: $\Delta s
\sim - (\alpha / 2 \pi) \times \Delta G$, which would require $\Delta G
\gappeq 2$~
\footnote{However, this suggestion is controversial, in particular
because the gluon renormalization of the quark spin is
scheme-dependent~\cite{manjaff}.}.
Thus there are two conceptually interesting values for $\Delta G$: a value
$\sim 1/2$ which would be 
comparable to the proton spin and hence make an 
important direct contribution to it, and a value $\sim 2$ which could 
`explain away' the strange quark spin.
It is worth keeping in mind that both of these effects require positive
$\Delta G$, and that very few models suggest a negative value.

First attempts to estimate $\Delta G$ from data have been made using
measurements of the polarized structure function $g_1$ at different
momentum scales $Q^2$ to extract $\Delta G$ indirectly using
next-to-leading order (NLO) QCD~\cite{Blumlein:2002be,Hirai:2003pm}. These
first attempts have been
inconclusive, unable to exclude any of the interesting theoretical
possibilities $\Delta G \sim 0, 0.5$ or 2. It is hoped that new
structure function measurements, e.g., by COMPASS~\cite{COMPASSprop}, will 
be able to refine this indirect extraction of $\Delta G$.

However, the main objective of COMPASS is the direct determination of
$\Delta G$ via asymmetries in the production of ${\bar c} c$ and
high-$p_T$ hadron pairs. The direct determination of
$\Delta G$ is also a key objective of the RHIC polarized beam
programme~\cite{pRHICprop}.  First results from both PHENIX at
RHIC~\cite{PHENIX} and COMPASS at both high~\cite{COMPASS} 
and low $Q^2$~\cite{COMPASSlowQ2} have recently been
announced, the former on the asymmetry $A_{LL}$, and the latter on the
asymmetry in high-$p_T$ hadron-pair production. The high-$Q^2$ COMPASS 
result was in fact the
third result on this asymmetry, having been preceded by measurements by
the HERMES~\cite{HERMES} and SMC~\cite{SMC} collaborations.

The current PHENIX result is difficult to analyze in terms of $\Delta
G$, since the error is still quite large, the central value is
outside the physical region, and the relation to $\Delta G$ is not 
single-valued. The errors in the high-$Q^2$ measurement
by COMPASS are no smaller than those of the previous HERMES and SMC
measurements, but the errors in their low-$Q^2$ are significantly 
smaller, and COMPASS benefits in the comparison with HERMES from
having more generous kinematics, which reduces potential issues related
to higher-twist effects. In fact, as we discuss in more detail below,
the COMPASS, SMC and HERMES data seem to be telling a consistent story.

In this paper we make a preliminary combination of the
HERMES~\cite{HERMES}, SMC~\cite{SMC} and
COMPASS~\cite{COMPASS,COMPASSlowQ2} data, seeking a first direct
indication of the possible magnitude of $\Delta G$ based on
hadron-asymmetry data.  Despite their limited precision, these data are
already precise enough for such an analysis to be carried out within the
framework of existing parametrizations of the possible polarized gluon
distribution.  On the other hand, they are not yet sufficiently precise to
merit a fully-fledged NLO fit.

We find that $\Delta G$ is unlikely to be
as large as was desired in attempts to 'explain away' the negative value
of $\Delta s$, though a substantial gluonic contribution to the spin of
the proton can certainly not be excluded. Using three different
parametrizations of the polarized gluon distribution, we find central
values of $\Delta G \sim 0.5$, with errors of similar magnitude. The
suggestion that $\Delta G \ge 2$ is disfavoured by $\Delta \chi^2 \sim 9$
to 20, depending on the parametrization adopted, assuming that the
polarized gluon distribution does not exhibit unexpected behaviour outside
the limited $x$ range covered by the current experiments.

\section{Available Experimental Information}

We first review the relevant experimental information that is
currently available. Three recent attempts to extract
$\Delta G$ from NLO analyses of deep-inelastic structure function
data yield the following estimates:
 \beq
\begin{array}{cccrcr}
\Delta G \, & = & \, 1.026 \pm 0.549
 &\hbox{Set 3, Ref.~\cite{Blumlein:2002be}}&\equiv& {\rm BB3}
\\
 \Delta G \, & = & \, 0.931 \pm 0.669
 &\qquad\qquad\hbox{Set 4, Ref.~\cite{Blumlein:2002be}} &\equiv &{\rm
BB4}\\
\Delta G \, & = & \, 0.533 \pm 1.931
 &\hbox{Ref.~\cite{Hirai:2003pm}} &\equiv& {\rm AAC}
\end{array}
\label{indirect}
\eeq
 The first two estimates were made by
the same group, and the difference between their central values
may be indicative of the systematic errors in this indirect
approach to $\Delta G$ that are associated with the choice of
parametrization of the polarized gluon distribution. The larger
error in the third estimate may mark a more realistic assessment
of the systematic errors in this indirect approach. According to these 
analyses, each of 
the parametrizations used in~\cite{Blumlein:2002be,Hirai:2003pm} would be
compatible with $\Delta G = 2$ at the 2-$\sigma$ level.

The measurements of the high-$p_T$ hadron production
asymmetry that we use here are the following:
 \beq
  \begin{array}{lcc}
 {\rm HERMES~\cite{HERMES}:}&
 \Delta G/G = \phantom{{-}}0.41 \pm 0.18 \pm 0.03 \qquad &0.06 < x_G <
0.28,\\
\hfill\\
 {\rm SMC~\cite{SMC}:} \;& \Delta G/G
  =   {-} 0.20 \pm 0.28 \pm 0.10 \qquad
&\langle x_G\rangle = 0.07,\\
\hfill\\
{\rm COMPASS}~\cite{COMPASS}:  \; &\Delta G/G  =
\phantom{{-}}0.06 \pm 0.31 \pm 0.06 \qquad &\langle x_G \rangle  =
0.13,
\\
&&Q^2 > \hbox{1  GeV}^2 
\\
\hfill\\
{\rm COMPASS}~\cite{COMPASSlowQ2}:  \; &\Delta G/G  =
\phantom{{-}}0.024 \pm 0.089 \pm 0.057 \qquad &\langle x_G \rangle  =
0.095.
\\
&&Q^2 < \hbox{1  GeV}^2
\\
\hfill\\
 \end{array}
 \label{direct}
  \eeq
where we have
indicated in each case the available information on the kinematic range of 
the measurement. 

In addition, COMPASS has recently released preliminary results 
\cite{COMPASS_open_charm} for gluon polarization from 
open charm, based on the 2002-2003 data:
\hfill\break
$\Delta G/G = {-}1.08 \pm 0.73$ at $\langle x_G =0.15 \rangle$,
RMS=0.08.
This channel has very little background, but very low statistics, resulting
in large errors. We therefore do not use these preliminary open charm 
data in our fits, but note that that the central value of $\Delta G$ is
negative, providing additional qualitative evidence against a large
positive $\Delta G$.

In
order to convert the measurements \eqref{direct}
into estimates of $\Delta G$,
one needs to assume a suitable form for the unpolarized gluon
distribution $G(x,Q^2)$ and specify the relevant momentum transfer
scale $Q^2$. As our defaults, we use a recent MRST
\cite{Martin:2004ir} gluon distribution and assume that $Q^2 \sim
5$~GeV$^2$, and we discuss later the sensitivity to the assumed
value of $Q^2$.

\section{Fits to Asymmetry Data}

In making our direct estimates of $Q^2$, we assume three trial
forms for the polarized gluon distribution $\Delta G(x,Q^2)$, proposed
by the 
groups mentioned earlier~\cite{Blumlein:2002be,Hirai:2003pm}. The explicit
expressions for the three parametrizations are available as
FORTRAN codes that can be downloaded from the HEPDATA site~\cite{HEPDATA}.
Schematically, they can be written in the
following form which emphasizes the overall normalization:
 \beq
\begin{array}{lccr}
 {\rm BB3:}  \qquad \;& \Delta G(x, Q^2) \, & = & \, A_{BB3}\cdot
f_{BB3}(x,Q^2)\,\cite{Blumlein:2002be},\\
 {\rm BB4:}   \;& \Delta G(x, Q^2) \, & = & \, A_{BB4}\cdot
f_{BB4}(x,Q^2)\,\cite{Blumlein:2002be},\\
 {\rm AAC:}   \;& \Delta G(x, Q^2) \, & = & \, A_{AAC}\cdot
f_{AAC}(x,Q^2)\,\cite{Hirai:2003pm}.
 \end{array}
 \label{params}
 \eeq
We treat the overall normalizations $A_{ijk}$ as free parameters, but
retain as defaults the values of the other parameters chosen 
in~\cite{Blumlein:2002be} and the central fit values found 
in~\cite{Hirai:2003pm}.
We then fit the overall normalizations to the three asymmetries
(\ref{direct}), and hence determine the integrated gluon polarization
$\Delta G$. 

\begin{figure}[t]
\strut\kern3em
\epsfig{figure=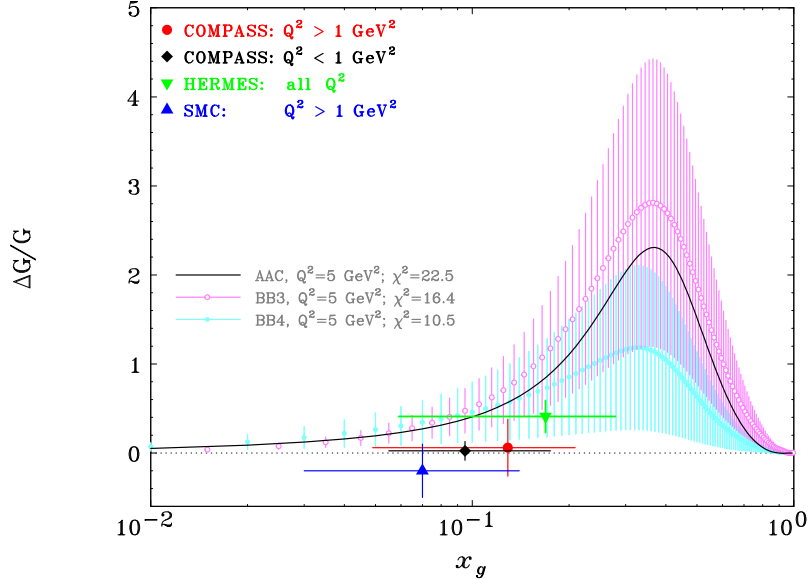,angle=90,width=0.78\textwidth}
\strut\vskip-0.4cm
\caption{\it Experimental results for $\Delta
G/G$~\protect\cite{COMPASS,COMPASSlowQ2,HERMES,SMC} compared with the
theoretical parametrizations
(\ref{params})~\protect\cite{Blumlein:2002be,Hirai:2003pm}, using
normalizations adjusted to yield $\Delta G = 2$.}
 \label{DeltaGfigA}
\end{figure}

The limited precision of the presently available
asymmetry data 
is insufficient to merit a fully-fledged multi-parameter NLO fit. 
Even with the much more precise deep-inelastic scattering data
set, the authors of Ref.~\cite{Blumlein:2002be}
found it necessary to to give up on a fully-fledged fit to the
parameters of the form they proposed for $\Delta G(x,Q^2)$, imposing by 
hand several
constraints and leaving the overall normalization as the main free 
parameter. On the other hand,
Ref.~\cite{Hirai:2003pm} did not impose supplementary constraints on their 
parametrization of the polarized gluon
distribution, which is why the errors for $\Delta G$ that they quote are 
considerably larger.

\strut\vskip-0.9cm
Our procedure is clearly hostage to unforeseen properties
of the polarized gluon distribution $\Delta G(x, Q^2)$ at values of $x$
outside the experimental ranges given in (\ref{direct}). This
possibility would introduce a systematic error that we are unable to
quantify.

\begin{figure}[t]
\strut\kern3em
\epsfig{figure=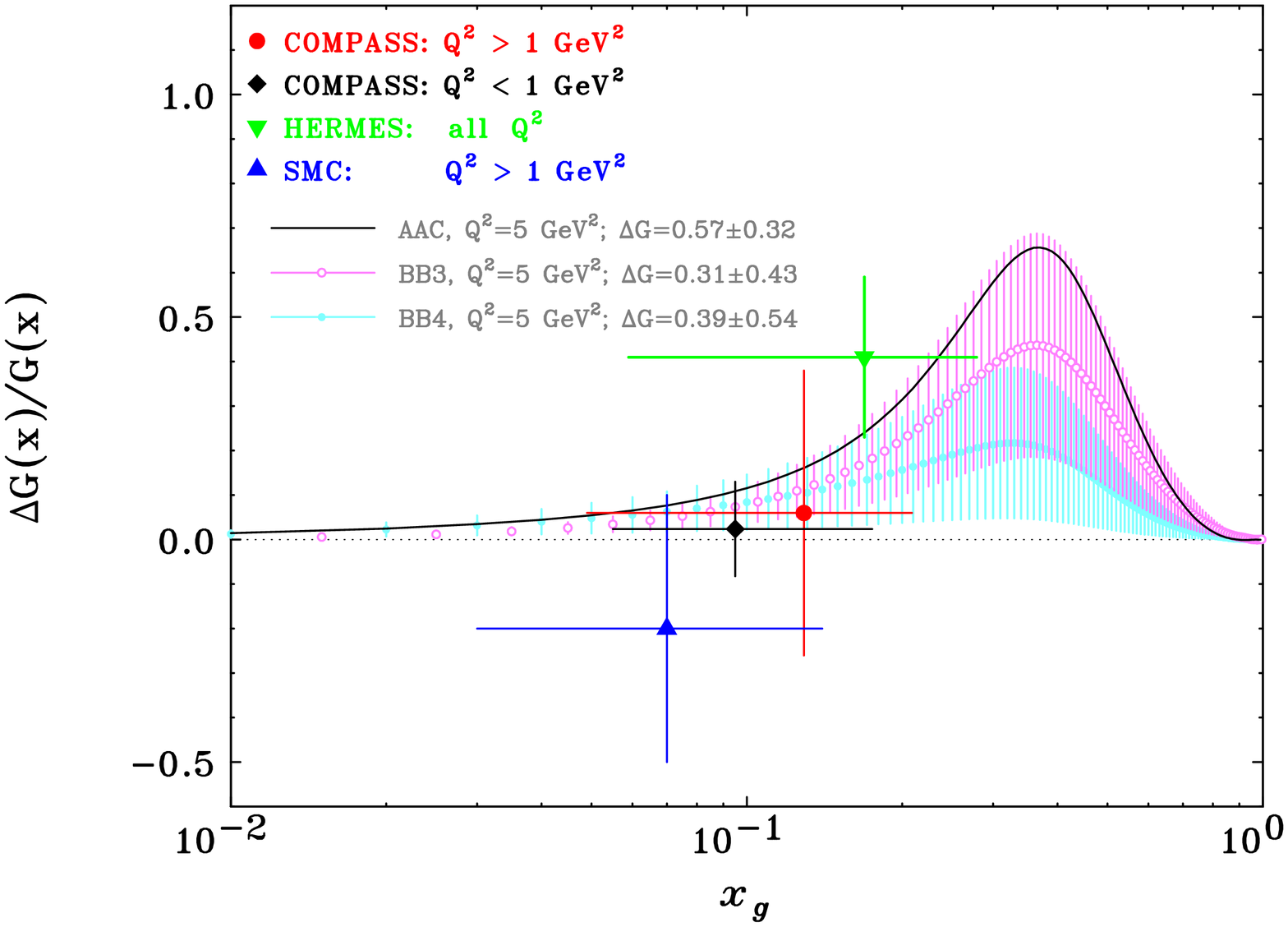,angle=90,width=0.8\textwidth}
\caption{\it Experimental results for $\Delta
G/G$~\protect\cite{HERMES,SMC,COMPASS,COMPASSlowQ2} compared with the
theoretical parametrizations
(\ref{params})~\protect\cite{Blumlein:2002be,Hirai:2003pm}, with
normalizations adjusted to yield best fits to the data.}
\label{DeltaGfigB}
\end{figure}

\strut\vskip-0.9cm
Before fitting the values of the normalizations $A_{ijk}$ in the different
para\-me\-tri\-zations, we have first assumed values that correspond to an
integral $\Delta G = 2$~\footnote{We repeat that such a value was 
compatible with the NLO analyses of \cite{Blumlein:2002be,Hirai:2003pm}, 
though not the central value suggested by their fits.}, and assessed 
the goodness of fit by evaluating the corresponding $\chi^2$ functions. 
As seen in Fig.~\ref{DeltaGfigA},
each of the parametrizations (\ref{params}) reproduces the general trend
of the values of $\Delta G$ indicated by the data from the SMC, COMPASS
and HERMES. However, each of the parametrizations also violates the
unitarity bound $\Delta G (x)  \le G (x)$ at larger $x$. Ignoring this
problem for the time being, we find $\chi^2 = 16.4,\ 10.5$ and 22.5 for the
BB3, BB4 and AAC functions, respectively. If one caps each 
polarized distribution by the unitarity bound, keeping the same 
normalization at lower $x$ where there are measurements, the integrals 
are reduced to $\Delta G = 1.63, 1.99$ and 1.80, respectively. If one 
were to attempt to compensate for these reductions in $\Delta G$ by 
increasing the normalization factors $A_{BB3, BB4, AAC}$, the $\chi^2$ 
values would each be increased.

Much better fits to the parametrizations (\ref{params}) are obtained when
the overall normalizations are allowed to float: $\chi^2 = 1.1,\ 1.5$ and
2.5\, and, as seen in Fig.~\ref{DeltaGfigB}, each parametrization now
respects the unitarity bound $\Delta G(x) \le G(x)$ for all $x$. In view
of all the assumptions and uncertainties, we are reluctant to quote the
differences in $\chi^2$, $\Delta \chi^2 = 15.3,\ 9.0$ and 20.0\,, as numbers
of standard deviations by which $\Delta G \ge 2$ is disfavoured. 
In particular, the $\Delta \chi^2$ found for the AAC fit would 
change significantly if the full freedom of the parametrization 
were explored. However,
it is clear that the current asymmetry data offer no hint in favour of
the option that $\Delta G \ge 2$. The following are the best fit values of 
$\Delta G$ that we
find for each of the parametrizations (\ref{params}) and the formal
errors:

\vbox{
\begin{eqnarray}
{\rm BB3:}  \; \Delta G \, & = & \, 0.31 \pm 0.43,\qquad \chi^2=1.1 \,,\\
\nonumber {\rm BB4:} \; \Delta G \, & = & \, 0.39 \pm 0.54,\qquad
\chi^2=1.5 \,,\\
\nonumber {\rm AAC:}  \; \Delta G \, & = & \, 0.57 \pm 0.32\qquad
\,\,\,\chi^2 =2.5\,. \label{bestfits}
\end{eqnarray}}
\noindent
The best fits are compared with the data in Fig.~\ref{DeltaGfigB}.

In the
preceding discussion we have compared the experimental data with
polarized gluon parametrizations $\Delta G(x,Q^2)$ evaluated at a
canonical value $Q^2=5$ GeV$^2$. However, each of the measured
values of $\Delta G/G$ in eq.~(\ref{direct}) results from
averaging by the relevant experiments over a  range of values of
$x$ and $Q^2$. The limited information in the 
experimental papers and the low precision of the currently available
data make it impossible to provide an accurate estimate of
systematic errors due to this averaging. We can, however, get a
semi-quantitative estimate of the relevant error by repeating the
fitting procedure at several values of $Q^2$, as shown in Table I.

\def\tstrut{\vrule height 1.1em depth 0.5em width 0pt}
\def\bb{\kern-0.3em}
\vbox{
\begin{center}
Table I
\\
\hfill\\
\strut\kern-0.5em
\begin{tabular}{|r||c|c|c||c|c|c||c|c|c|} \hline
& \multicolumn{3}{c||}{BB4}
& \multicolumn{3}{c||}{BB3}
& \multicolumn{3}{c |}{AAC}
\\ \cline{2-10}
\tstrut $Q^2$\,\,\,\,
&          &$\chi^2$&$\chi^2$
&          &$\chi^2$&$\chi^2$
&          &$\chi^2$&$\chi^2$
\\ 
\tstrut\kern-0.3em$\hbox{GeV}^2$\kern-0.4em\tstrut
   &$\Delta G$& \kern-0.4em best \kern-0.9em   &   for
   &$\Delta G$& \kern-0.4em best \kern-0.9em   &   for
   &$\Delta G$& \kern-0.4em best \kern-0.9em   &   for
\\
   &          & fit    &  \kern-0.4em $\Delta G{=}2$ \kern-0.7em
   &          & fit    &  \kern-0.4em $\Delta G{=}2$ \kern-0.7em
   &          & fit    &  \kern-0.4em $\Delta G{=}2$ \kern-0.7em
\\ \hline
\hline
 1.5 &\bb$0.31{\pm}0.40$\bb& 1.6 &19.7 &\bb$0.46{\pm}0.44$\bb&1.0&13.4&\bb $0.58{\pm}0.31$\bb& 2.2& 22.8
\\ \hline
 2.0 &\bb$0.33{\pm}0.43$\bb& 1.5 &16.5 &\bb$0.33{\pm}0.37$\bb&1.2&21.2&\bb $0.57{\pm}0.31$\bb& 2.3& 23.4
\\ \hline
 5.0 &\bb$0.39{\pm}0.54$\bb& 1.5 &10.5 &\bb$0.31{\pm}0.43$\bb&1.1& 16.4&\bb $0.57{\pm}0.32$\bb& 2.5&  22.5
\\ \hline
10.0 &\bb$0.43{\pm}0.62$\bb& 1.5 &6.5 &\bb$0.30{\pm}0.48$\bb&1.1& 13.5&\bb
$0.58{\pm}0.33$\bb& 2.6&  20.6
\\ \hline
\hline
\end{tabular}
\end{center}
{\it \small Fits to the HERMES, SMC and COMPASS $\Delta G/G$
data (\ref{direct}), 
for $Q^2=1.5$, 2, 5 and 10 GeV$^2$, using the parametrizations (\ref{params})
 of 
$G(x,Q^2)$. 
For each parametrization we list the best-fit value of $\Delta G$ and
its $\chi^2$, as well as
the $\chi^2$ value corresponding to $\Delta G=2$.}
}
\hfill\break
The trend shown by these fits is clear: for $1.5 \le Q^2 \le 10$ GeV$^2$
(which includes the preferred value $Q^2 = 3$~GeV$^2$ quoted 
in~\cite{COMPASSlowQ2} and the three parametrizations (\ref{params}),
the best fit values $\Delta G$ range between $0.30\pm0.48$ and 
$0.58\pm0.33$. In all cases the $\Delta G=2$ value is significantly
disfavoured.

The current discussion is based on the COMPASS 2002 - 2003 data with $Q^2
> 1$ GeV$^2$ \cite{COMPASS} and
approximately 10 times more data with $Q^2 < 1$ GeV$^2$
\cite{COMPASSlowQ2}.
The present statistics will be approximately doubled with
the 2004 data, 
When the full 2004 set of all $Q^2$ COMPASS data is analyzed,
the statistical error on $\Delta G/G$ is expected to go down to
$\pm0.05$~\cite{COMPASS}, compared with $\pm 0.31$ in the currently
available 2002-03 data set. However, the inclusion of $Q^2 < 1$ GeV$^2$
data introduces additional theoretical uncertainties.

Still, it is clear that the inclusion of the full 2004 data set 
will significantly increase the precision with which the first moment
$\Delta G$ can be estimated.
To see the effect of the increased precision of the future COMPASS data, 
we have repeated the current fits, setting the
COMPASS $\Delta G/G$ error at $\pm 0.06$, while keeping the HERMES and 
SMC data unchanged. 
The expected error on $\Delta G$ shrinks down to 
$\pm0.44$, $\pm0.55$ and $\pm0.16$, for the BB3, BB4 and AAC 
parametrizations, respectively.
In the cases of the two BB parametrizations, the expected 
error reduction is rather modest, but we
attribute this to the fact that in this exercise 
the fit includes only one $x$ value with increased precision.
This underlines the importance
of providing the high-precision 
values of $\Delta G/G$ over a range of $x$.

\section{Summary and Prospects}

We have made in this paper a first direct estimate of the net
gluon polarization in the nucleon, based on hadron-asymmetry data
in deep-inelastic
scattering~\cite{COMPASS,COMPASSlowQ2,HERMES,SMC}. Despite
being very crude and incomplete in its kinematic coverage, this
direct estimate has an error that is comparable with that provided
indirectly by NLO analyses of deep-inelastic structure
functions~\cite{Blumlein:2002be,Hirai:2003pm}. We find a favoured value
of $ \Delta G \sim 0.5$, with a formal error of similar magnitude.
Values of $ \Delta G \ge 2$ are disfavoured by $\Delta \chi^2 \sim
9$ to 20, depending on the parametrization of the polarized-gluon
distributions that is used.

There are good prospects for a significant improvement soon in the
accuracy with which $\Delta G$ is known, thanks to new data from
COMPASS and RHIC.  The present data are insufficient to exclude
strongly the hypothesis that all the apparent negative value of
$\Delta s$ might be induced by gluons via renormalization in one
particular scheme. The forthcoming data should be able to resolve
this issue. However, they might not be able to determine whether
gluons carry a large part of the nucleon spin, $ \Delta G \sim
1/2$, or whether their contribution is as small as that due to the
quarks, as expected qualitatively in chiral soliton 
models~\cite{BEKEK}. There
are surely still many interesting twists and turns still to come
in our understanding of the nucleon spin, but direct
determinations of the gluon spin now seem poised to make an
important step forward.

\section*{Acknowledgements}

The research of one of us (M.K.) was supported in part by a grant from the
Israel Science Foundation administered by the Israel
Academy of Sciences and Humanities.
We thank 
Elke Aschenauer,
Colin Bernet,
Catarina Quintans,
Ewa Rondio,
Naohito Saito
and
Robert Thorne
for discussions and correspondence.

\end{document}